\documentstyle[12pt,colordvi,epsf]{article}

\begin{document}
\renewcommand{\thefootnote}{\fnsymbol{footnote}}
\begin{center}
{\LARGE ON THE NON-ORTHONORMALITY OF LIPPMANN-SCHWINGER-LOW
STATES}\vspace{.42in}\\
{\large V. J. Menon$^1$ and B. K. Patra$^2$}\vspace{.2in}\\
{\it $^1$ Department of Physics, Banaras Hindu University, Varanasi 221 005,
India\\
$^2$ Variable Energy Cyclotron Centre, 1/AF Bidhan Nagar, Calcutta 700 064,
India\\}
\vspace{.42in}
\underline{Abstract}
\end{center}

It is pointed out that for a general short-ranged potential the
Lippmann-Schwinger-Low scattering state $|\psi^L_k \rangle$ does not
strictly satisfy the Schrodinger eigen equation, and the pair
$|\psi^L_n\rangle$, $|\psi^L_k\rangle$ is mutually nonorthogonal
if $E_n=E_k$. For this purpose, we carefully use an infinitesimal adiabatic
parameter $\epsilon$, a nonlinear relation among  transition amplitudes, and
a separable interaction as illustration. \vspace{.83in}\\


\pagebreak
{\Large \bf \underline {Introduction}} \\

The Lippmann-Schwinger-Low (LSL) integral equations for state vectors
and transition matrices form the backbone of quantum scattering theory [1].
They provide the basis for deriving the Born series in wave mechanics [2], 
reaction amplitudes in rearrangement collisions [3], Dyson's perturbation 
expansion
in the Dirac picture [4], and various cross sections in old-fashioned quantum
electrodynamics [5]. The aim of the present paper is to examine some features of the LSL equations which have not been treated adequately in the existing 
literature. To be more precise, Lemmas A, B, C and D below answer the following
four questions : (i) Are the LSL representations strictly equivalent to the
underlying Schrodinger eigen equations? (ii) What is a general off/on 
energy-shell unitarity-like relation obeyed by the LSL transition amplitudes? 
(iii) Do various LSL state vectors accurately satisfy the orthonormality 
relations mentioned by Goldberger-Watson [6]? (iv) Can we confirm the
results explicitly in the case of a separable potential for
which the LSL solutions can be obtained in closed form [7]?\\

{\large \bf \underline {Preliminaries}}\\
We denote the free and full Hamiltonian operators by $H^o$ and $H \equiv H^o +
V$ respectively with $V$ being a short-range interaction. Their continuum
eigenkets obey the Schrodinger (superscript S) equations
\begin{eqnarray}
(E_k -H^o)|k \rangle &=&0 \\
(E_k - H) |\psi_k^S \rangle &=& 0
\end{eqnarray}
where the masses are assumed to be renormalized so that energies do not shift.
For later convenience we also introduce the free resolvent $G_k^o$, the complex
projector $\eta_k^o$ onto free states of energy $E_k$, $\pi$ times a Dirac delta
$D_k^o$, related functions $\mu_{nk}$ and $d_{nk}$ along with a useful
identity via
\begin{eqnarray}
G_k^o = \frac{1}{E_k-H^o+i\epsilon}~;~\eta_k^o=i\epsilon G_k^o~;~\mu_{nk}
=\frac{i\epsilon}{E_k-E_n+i\epsilon}
\end{eqnarray}
\begin{eqnarray}
D_k^o = \pi \delta (E_k-H^o) = \epsilon G_k^{o\dag} G_k^o~;~ d_{nk} = \frac{\epsilon}
{{(E_k-E_n)}^2 + \epsilon^2}
\end{eqnarray}
\begin{eqnarray}
\frac{G_n^{o\dag}-G_k^o}{E_k-E_n+i\epsilon} = \frac{E_k-E_n +2i\epsilon}{E_k-E_n+
i\epsilon}~G_n^{o\dag} G_k^o = \left( 1 +\mu_{nk} \right) G_n^{o \dag} G_k^o
\end{eqnarray}
where $\epsilon \rightarrow +0$ is an adiabatic parameter, and $\mu_{nk}$
and $d_{nk}$ vanish if $E_n \neq E_k$. It is customary to replace Eq.(1) $\&$ 
(2) by the LSL representations (labeled by the superscript $L$)
\begin{eqnarray}
|\psi^L_k \rangle &=& |k\rangle + G_k^oV|\psi^L_k \rangle~~~~~~~~~~~~~~~~~~~~~:~LS \\
&=& |k \rangle + {(E_k-H+i\epsilon)}^{-1}V |k \rangle~~~~~~:~Low
\end{eqnarray}
obeying plane $+$ outgoing boundary conditions. Our objective is to
propose a few Lemmas on some algebraic properties of $|\psi^L_k \rangle$ below
by paying careful attention to the $\epsilon$ factors.\\

{\large \bf \underline {LEMMA A (COMPARISON WITH SCHRODINGER)~:}}\\

``In sharp contrast to the underlying Eq.(1) the LSL states satisfy
\begin{eqnarray}
(E_k-H+i\epsilon) |\psi^L_k \rangle &=& i \epsilon | k \rangle~,
\end{eqnarray}
or equivalently
\begin{eqnarray}
(E_k-H) |\psi^L_k \rangle &=& -\eta_k^o V |\psi^L_k\rangle~" 
\end{eqnarray}

{\bf \underline {Proof}}\\
Eq.(8) follows from the application of the operator $(E_k-H^o+i\epsilon)$
on Eq.(6) or $(E_k-H+i\epsilon)$ on Eq.(7). It suggests that $|\psi^L_k
\rangle$ is not a strict eigenket of $H$ for any nonzero infinitesimal 
$\epsilon$. Eq.(9) is an outcome of the fact that $i \epsilon |\psi^L_k \rangle=
\eta_k^oV |\psi^L_k \rangle$ is generally a nonzero ket. Indeed, the matrix element
$\langle n|\eta_k^oV|\psi_k^L \rangle = \mu_{nk} \langle n|V|\psi^L_k \rangle$ becomes the
on-shell transition amplitude if $E_n=E_k$.\\

{\large \bf \underline {LEMMA B (NONLINEAR RELATION FOR T-MATRIX)~:}}\\

``The amplitudes $T^L_{nk} \equiv \langle n|V|\psi^L_k \rangle$ fulfill a nonlinear relation
\begin{eqnarray}
(T_{nk}^L - T^{L^\ast}_{kn})/(E_k-E_n +i\epsilon) = - (1+\mu_{nk}) C^L_{nk} \\ 
C^L_{nk} = \langle \psi^L_n|V G_n^{o\dag} G_k^o V | \psi_k \rangle~"
\end{eqnarray} 

{\bf \underline {Proof}}\\
From Eq.(6) we first obtain $ \langle n|$ and thereby write
\begin{eqnarray}
T^L_{nk} =  \langle \psi_n^L|V|\psi_k^L \rangle  - \langle \psi_n^L|VG_n^{o\dag}V|
\psi^L_k \rangle
\end{eqnarray}
Subtracting a similar expression for $T^{L^\ast}_{kn} \equiv \langle \psi^L_n|V|k \rangle$ 
and
employing the identity (5) the desired Lemma follows.\\
Incidentally, in the special case of $E_n=E_k$ our Eqs.(10), (11) reduce to the usual
on-shell unitarity relation [2-5] viz.
\begin{eqnarray}
\left[ T^L_{nk} - T^{L^\ast}_{kn} \right]_{E_n=E_k} = -2i A^L_{nk} \\
A^L_{nk} = {\left[ \epsilon C^L_{nk} \right]}_{E_n=E_k} = \langle \psi^L_n|V D^o_k V |\psi^L_k \rangle
\end{eqnarray}
\vspace{1in}\\

{\large \bf \underline {LEMMA C (NONORTHONORMALITY)~:}} \\

``Consider the overlap $I^L_{nk} \equiv \langle \psi^L_n|\psi^L_k \rangle$ between two arbitrary outgoing
LSL states. In sharp contrast to the conventional erroneous value [6] $\langle 
n|k \rangle$
for the overlap its correct value is
\begin{eqnarray}
I^L_{nk} = \langle n|k \rangle - d_{nk} A^L_{nk}~"
\end{eqnarray}
with $d_{nk}$ given by Eq.(4) and $A^L_{nk}$ by Eq.(14).\\

{\bf \underline {Proof}}\\
Upon using the Low form for $\langle \psi^L_n|$ and the LS form for $|\psi^L_k
\rangle$ 
(cfs. Eqs. 6,7) one finds
\begin{eqnarray}
\langle \psi^L_n |\psi^L_k \rangle = \langle n| \psi^L_k \rangle + \langle 
n|V {(E_n-H-i\epsilon)}^{-1} |\psi^L_k \rangle
\end{eqnarray}
In the usual Goldberger-Watson treatment (labeled by the superscript G) one 
erroneously assumes that $H |\psi^L_k \rangle = E_k |\psi^L_k \rangle $ and
reduces Eq.(16) into [6]
\begin{eqnarray}
I^G_{nk} = \langle n|k \rangle + \langle n|V \left( \frac{1}{E_k-E_n+i\epsilon} + \frac{1}
{E_n-E_k-i\epsilon} \right) |\psi^L_k \rangle = \langle n|k \rangle
\end{eqnarray}
In our opinion the use of Eqs. (8), (9) as eigenket statement is quite risky
and it is much safer to employ the LS representations (6) for both $\langle
\psi^L_n|$
and $|\psi^L_k \rangle $. Then
\begin{eqnarray}
I^L_{nk} &=& \langle n|k \rangle + \langle n|G_k^o V |\psi^L_k \rangle + 
\langle \psi^L_n |V G_n^{o\dag} |k \rangle 
\nonumber\\
& & + \langle \psi^L_n |V G_n^{o\dag} G_k^o V |\psi^L_k \rangle
\end{eqnarray}
which is readily shown to coincide with the Lemma (15) in view of the properties
(Eq.(10)) and (Eq.(14)). The fact that $I^L_{nk}$ reduces to $\langle n|k \rangle$ if $E_n \neq E_k$ but fails
to do so if $E_n=E_k$ is very disturbing because it implies that the set of
LSL states $|\psi^L_n \rangle$ which are degenerate at a given collision energy $E_k$
are mutually nonorthogonal.\\

{\large \bf \underline {LEMMA D (ILLUSTRATION)~: }}\\
``Consider a rank $1$ separable potential [7] $V= \lambda |g \rangle
\langle g|$ with $\lambda$
being a real coupling and $|g \rangle$ a wave packet. Then, the overlap 
$ \langle \psi^L_n
|\psi^L_k \rangle$ can be independently shown to be
\begin{eqnarray}
I^L_{nk} = \langle n|k \rangle - d_{nk} {\lambda}^2 g_n g_k^\ast~\frac{ \langle g|D_k^o|g
\rangle }{  
 \Delta_n^{\ast} \Delta_k}
\end{eqnarray}
where the form factor $g_k$ and Fredholm determinant $\Delta_k$ are
defined by
\begin{eqnarray}
g_k= \langle k|g \rangle~;~ \Delta_k = 1 - \lambda \langle g|G_k^o|g 
\rangle~"
\end{eqnarray}

{\bf \underline{Proof}}\\
With $V=\lambda|g \rangle \langle g|$, Eq.(6) is readily solved in closed form as
\begin{eqnarray}
|\psi^L_k \rangle  &=& |k \rangle + G_k^o |g \rangle \left( \lambda g_k^\ast/\Delta_k \right) 
\nonumber\\
\langle \psi^L_n| &=& \langle n| + \left( \lambda g_n/\Delta_n^\ast \right) \langle g|G_n^{o\dag}
\end{eqnarray}
Then it is straightforward to compute
\begin{eqnarray}
I^L_{nk} =  \langle n|k \rangle - \frac{\lambda^2 g_n g_k^\ast}{\Delta_n^\ast \Delta_k}
\left[ \frac{\Delta_k - \Delta_n^\ast}{\lambda(E_k-E_n+i\epsilon)} -
 \langle g|G_n^{o\dag}G_k^o|g \rangle \right]
\end{eqnarray}
which coincides with the stated lemma in view of the useful identity
(5). Of course, the illustrative Eq.(19) and the general result Eq.(15)
are in complete agreement although they were derived by different methods.\\

{\Large \bf \underline {CONCLUSIONS}}\\
The main findings of the present paper are contained in Lemmas A, B, C, and D. 
The nonorthogonality of the LSL states (for $E_n=E_k$, $n \neq k$) implies
that, even in the absence of bound states, the Moller operator
connecting $|k \rangle$ to $|\psi^L_k \rangle$ may be nonunitary and $\sum_k | \psi^L_k
\rangle \langle
\psi^L_k|$ may loose its interpretation as the unit matrix. Several
standard results of scattering perturbation theory [1-7] based on the
LSL states may require re-examination. Before ending, it may
be added that the present work is not concerned with another
peculiarity of the LS equation - the Faddeev ambiguity [8] - arising
from the noncompactness of the kernel. We also believe that the 
time-dependence of the LSL states will be much richer than the
standard Schrodinger kets $|\psi_k^S (t) \rangle$ but this
aspect will be dealt-with in a future communication.

\noindent {\bf ACKNOWLEDGEMENTS} : 
VJM thanks the UGC, Goverment of India, New Delhi for financial support.

\pagebreak


\begin{thebibliography}{39}
\bibitem{1} B. A. Lippmann and J. Schwinger, Phys. Rev. {\bf 79}, 469 (1950).\\
F. E. Low, Phys. Rev. {\bf 97}, 1392 (1955). 
\bibitem{2} Charles J. Joachain, Quantum Collision Theory, 
North-Holland Publishing Company (Amsterdam), Lippmann-Schwinger
Equation and the Born series, pp. 161-162, 305-307, 409-410. 
\bibitem{3} M. L. Goldberger and K. M. Watson, Collision Theory, John
Wiley $\&$ Sons, Inc. (1964), L-S-L Equation, pp 197-199, Rearrangement 
Reactions, page 157. 
\bibitem{4} Silvan S. Schweber, An Introduction to Relativistic Quantum 
Field Theory, Harper $\&$ Row, Publishers, New York, N. Y., Lippmann
Schwinger equation pp.316, 327, $S$-matrix expansion (page 330) : 
$S~or~U(\infty,-\infty)=Pexp[-i\int_{-\infty}^{\infty}dt V_I(t)]=
1+ \sum_{n=1}^{\infty}{(-\frac{i}{\hbar})}^n \int dt_1...\int dt_n~P~V_I(t_1)...V_I(t_n)$  
\bibitem{5} Heitler, Quantum Theory of Radiation, page 132, 148. 
\bibitem{6} M.L. Goldberger and K. M. Watson, Collision Theory, $\langle
\psi_n | \psi_k \rangle = \delta_{nk}$, page 191, 199. 
\bibitem{7} Yoshio Yamaguchi, Phys. Rev. {\bf 95}, 1628 (1954).\\
C.S. Warke and R. K. Bhaduri, Nucl. Phys. {\bf A162}, 289 (1971).\\
B. Mulligan, L. G. Arnold, B Bagchi and T. O. Krause, Phys. Rev. {\bf C13}, 2131 (1976).
\bibitem{8} A. G. Sitenko, Lectures in Scattering Theory, pp.187-191.\\
Roger G. Newton, Scattering Theory of Waves and Particles, McGRAW-HILL
Book company (1966), Lippmann Schwinger Equation, pp.180-181, Low Equation
page 190, Faddeev Equation, page 556. 
\end{thebibliography}
\end{document}